\documentclass[aps,prl,reprint,groupedaddress]{revtex4-1}
\usepackage {graphicx}  
\usepackage{bm}        
\usepackage{amssymb}   

\begin{document}

\title{ Evidence for the Collective Nature of the Reentrant Integer Quantum Hall States of the Second Landau Level }

\author{N. Deng$^1$, A. Kumar$^1$,  M.J. Manfra$^{1,2}$ , L.N. Pfeiffer$^3$, K.W. West$^3$, 
        and G.A. Cs\'{a}thy$^1$ \footnote{gcsathy@purdue.edu}}

\affiliation{${}^1$ Department of Physics, Purdue University, West Lafayette, IN 47907, USA \\
${}^2$ Birck Nanotechnology Center,
School of Materials Engineering and School of Electrical and Computer Engineering,
Purdue University, West Lafayette, IN 47907, USA \\
${}^3$Department of Electrical Engineering, Princeton University, Princeton, NJ 08544\\}

\date{\today}

\begin{abstract}
We report an unexpected sharp peak in the temperature dependence of the magnetoresistance of the
reentrant integer quantum Hall states in the second Landau level. This peak defines the
onset temperature of these states. We find that in different spin branches
the onset temperatures of the reentrant states scale with the Coulomb energy. 
This scaling provides direct evidence that Coulomb interactions play an important role in the formation
of these reentrant states evincing their collective nature. 

\end{abstract}
\maketitle

The second Landau level (SLL) of the two-dimensional electron gas (2DEG) 
is astonishingly rich in novel ground states \cite{eisen02,xia04,csa10}. 
Recent experiments \cite{csa10,radu08,dolev08,willett10,vivek11,dolev11,csa11b}
suggest that there are both
conventional \cite{laughlin,jainCF} as well as exotic fractional quantum Hall states (FQHSs) 
\cite{moore91,greiter91} in this region. The study of the latter has
enriched quantum many-body physics with numerous novel concepts such as
paired composite fermion states with Pfaffian correlations, non-Abelian quasiparticles
\cite{moore91,greiter91,park98,read99,scarola00,rezayi00,read00,lu10}, 
topologically protected quantum computing \cite{dassarma05}, and established connections between the 2DEG 
and $p$-wave superconductivity in Sr$_2$RuO$_4$ and fermionic atomic condensates.

The eight reentrant integer quantum Hall states (RIQHSs) form another
set of prominent ground states in the SLL \cite{eisen02}.  
The transport signatures of the RIQHSs are consistent
with electron localization in the topmost energy level \cite{eisen02}. However, the nature of the localization
is not yet well understood. Depending on the relative importance of the electron-electron interactions,
the ground state can be either an Anderson insulator or a collectively pinned electron solid.

FQHSs owe their existence to the presence of the interelectronic Coulomb interactions \cite{laughlin,jainCF}. 
Since FQHSs and RIQHSs alternate in the SLL, it was argued that Coulomb interactions 
must be important and, therefore, the RIQHSs in the SLL must be electron solids \cite{eisen02}.
Subsequent density matrix renormalization group \cite{shibata03} and Hartree-Fock 
calculations \cite{goerbig03} also favored the electron solid picture
and predicted the solid phase similar to the Wigner crystal, but having one or more electrons 
in the nodes of the crystal \cite{goerbig03}.
Recently reported weak microwave resonances in one such RIQHS
are suggestive of but are far from being conclusive on the formation
of a collective insulator \cite{lewis05}.
Our understanding of the RIQHSs in the SLL, therefore, is still in its infancy
and the collective nature of these states has not yet been firmly established.
 
We report a feature in the temperature dependent magnetoresistance
unique to the the RIQHSs in the SLL, a feature which is used to define the onset temperature of these states.
The scaling of onset temperatures with the Coulomb energy reveals that Coulomb interactions 
play a central role in the formation of RIQHSs and, therefore, these reentrant states 
are exotic electronic solids rather than Anderson insulators. 
We also report an unexpected trend of the onset temperatures
within each spin branch. This trend is inconsistent with current theories
and can be understood as a result of a broken electron-hole symmetry. Explaining such
a broken symmetry of the RIQHSs is expected to impact our understanding of a similar
asymmetry of the exotic FQHSs of the SLL, including the one at $\nu=5/2$.

We performed magnetotransport measurements on a high quality GaAs/AlGaAs sample of density 
n=$3.0 \times 10^{11}$cm$^{-2}$ and of mobility $\mu $=$3.2 \times 10^{7}$cm$^{2}/$Vs. 
Earlier we reported the observation of a new FQHS at $\nu=2+6/13$ in this sample \cite{csa10}. 
The sample is immersed into a He-3 cell equipped with a quartz tuning fork viscometer used for 
$B$-field independent thermometry \cite{csa11}.

\begin{figure}[t]
 \includegraphics[width=1\columnwidth]{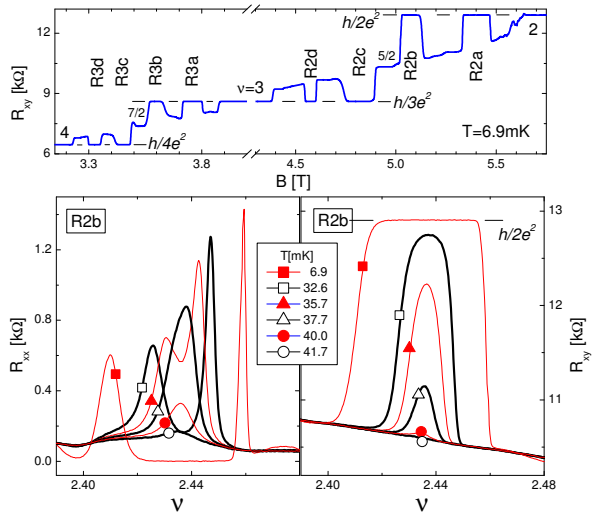}
 \caption{ The Hall resistance of the eight RIQHSs in the SLL at 6.9~mK (top panel) and 
 the temperature evolution of the RIQHS labeled $R2b$ (bottom panels). Numbers mark
 Landau level filling factors of importance.
 \label{Fig1}}
 \end{figure} 
 
In the top panel of Fig.\ref{Fig1} we show the Hall resistance $R_{xy}$ in the SLL at 6.9~mK. 
The data reveals numerous FQHSs and it is dominated by the eight RIQHSs. 
Starting with the states at the highest $B$-field 
we label the RIQHSs with $R2a$, $R2b$, $R2c$, and $R2d$ in the lower spin branch of the SLL (i.e. $2<\nu<3$) and with
$R3a$, $R3b$, $R3c$, and $R3d$ in the upper spin branch (i.e. $3<\nu<4$).
Here $\nu=nh/eB$ is the Landau level filling factor. 
RIQHSs have historically been predicted \cite{fogler96} and observed \cite{lilly99,du99} 
in high Landau levels (i.e. $\nu>4$). In contrast to the SLL, in high Landau levels there are only
four RIQHSs in each Landau level. 
These states develop at the lowest temperatures around non-integer filling factors and 
yet their Hall resistance $R_{xy}$ is quantized to a nearby integer plateau \cite{eisen02}. 

Because of the delicate nature of the RIQHS in the SLL 
\cite{eisen02,xia04,csa10,radu08,dolev11,csa11b,csa05,pan08,dean08,umansky09,nuebler10,xia10}
there is only scarce information available on their temperature dependence \cite{csa05,pan08,nuebler10}.
The lower panels of Fig.\ref{Fig1} show the details of the evolution of the 
longitudinal resistance $R_{xx}$ and $R_{xy}$ of $R2b$ with temperature $T$. 
The $R_{xx}(B)|_{T=6.9\text{mK}}$ curve has a wide zero flanked by two sharp spikes. 
As the temperature is raised, the spikes in $R_{xx}$ persist but they move closer to each other
and the width of the zero decreases.
At 32.6~mK $R_{xx}(B)$ does still exhibit the two spikes but instead of a zero it has a non-zero local minimum. 
The location in $B$-field of this minimum is $T$-independent and it defines the center $\nu_c=2.438$ of the $R2b$ state. 
At 35.7~mK the two spikes of $R_{xx}(B)$ have moved closer to each other and between them there is still a local minimum, 
albeit with a large resistance. A small increase in $T$
of only 2~mK leads to a qualitative change. Indeed, in contrast to curves at lower $T$,
$R_{xx}(B)|_{T=37.7\text{mK}}$ exhibits a single peak only. 
As the temperature is further raised, this single peak rapidly decreases until it merges into a low resistance background.
Simultaneously with the described changes of $R_{xx}$, $R_{xy}$ evolves from the quantized value $h/2e^2$ 
to its classical value $B/ne=h/\nu_c e^2$.

The behavior seen in Fig.1 can be better understood by measuring $T$-dependence at a fixed $\nu$.
In Fig.2 we show $R_{xy}$ versus $T$ near the center $\nu_c$ of the various RQIHSs. It is found that
$R_{xy}$ assumes the classical Hall resistance at high temperatures
and it is quantized to $h/2e^2$ or $h/3e^2$ at the lowest temperatures. 
Since 80\% of the change in $R_{xy}$ between these two values occurs over only 5~mK,
this change is very abrupt and it clearly separates the RIQHS at low $T$ from the classical gas at high $T$.
We interpret the inflextion point in $R_{xy}$ versus $T$ as being the onset temperature $T_c$ of the RIQHS. 
For reliable measurements in the vicinity of $T_c$ the temperature is swept slower than 10~mK/hour.

 \begin{figure}[b]
 \includegraphics[width=.9\columnwidth]{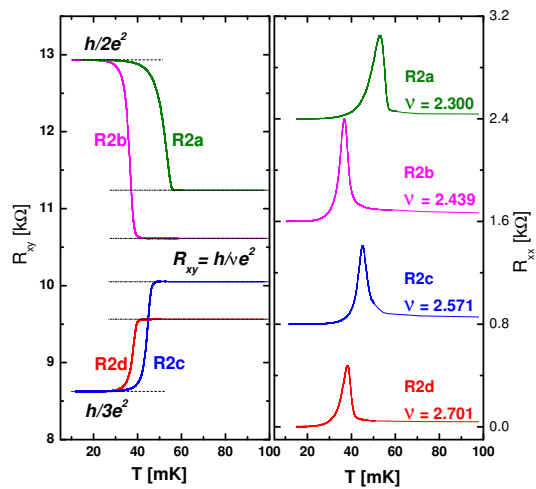}
 \caption{The evolution of the magnetoresistance of RIQHSs of the lower spin branch
 with temperature near the center  $\nu_c$ of each RIQHS. 
 For clarity, $R_{xx}(T)$ curves have been shifted vertically by 0.8~k$\Omega$.
 \label{Fig2}}
 \end{figure}
 
A transition from the classical Hall value to a quantized $R_{xy}$ with decreasing $T$ is observed not only
for the RIQHSs in the SLL but also in the vicinity of any developed integer or fractional quantum Hall state
and it is due to localization in the presence of a $B$-field. 
As seen in Fig.2, the $R_{xx}(T)|_{\nu=\text{fixed}}$ curves for the RIQHSs are non-zero
at high $T$, they vanish at low $T$, and they exhibit a sharp peak 
at the onset temperature $T_c$ defined above. In contrast, $R_{xx}(T)|_{\nu=\text{fixed}}$ 
of a quantum Hall state changes monotonically, without the presence of a peak. 
The sharp peak in $R_{xx}(T)|_{\nu=\text{fixed}}$ is, therefore, a signature of localization 
{\it unique} to the RIQHSs in the SLL and the peak temperature can be used as an alternative definition 
for the onset temperature $T_c$.

\begin{figure}[t]
 \includegraphics[width=.9\columnwidth]{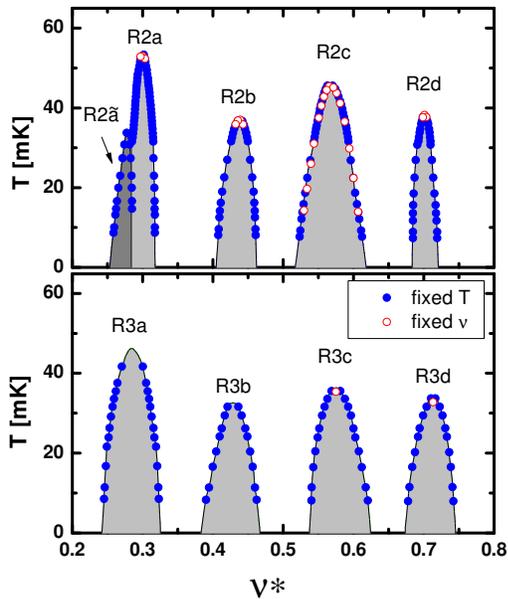}
 \caption{ 
 The phase boundaries of the eight RIQHS in the SLL in the  $\nu^*$-$T$ plane. The RIQHSs
 are stable within the shaded areas.
 Below 33~mK the $R2a$ state has a split-off state labeled $R2\tilde{a}$.
 \label{Fig3}}
 \end{figure}

Fig.3 represents the stability diagram of the RIQHSs in the $\nu^*$-$T$ plane. 
As described earlier, at a given $\nu$ the RIQHSs develop below the peak present
in the $R_{xx}(T)|_{\nu^*=\text{fixed}}$ curve. Such peaks are shown in Fig.2 for $\nu^* \approx \nu^*_c$, 
but similar peaks are also
present for nearby filling factors (not shown). Open symbols in Fig.3 are the peak temperatures $T_c$ 
as plotted against $\nu^*$. Similarly, the RIQHSs develop between the spikes of the
$R_{xx}(\nu)|_{T=\text{fixed}}$ curves. Such spikes are shown for $R2b$ in the lower panels of Fig.1.
The filling factors $\nu^*$ of the spikes for each RIQHS measured at a given temperature 
are marked with closed symbols in Fig.3. 
The excellent overlap of the two data sets in Fig.3 shows that the two definitions
used above self-consistently define the stability boundary of each RIQHS. The shaded areas 
within each boundary of Fig.3 represent the RIQHSs. FQHSs can develop only outside these shaded 
areas. The locations $\nu^*_{\text{high}}$ and $\nu^*_{\text{low}}$ of the spikes 
of the $R_{xx}(\nu)|_{T=\text{fixed}}$ curve measured at the lowest $T=6.9$~mK of our experiment
are listed in Table I.

We note that the $R2a$ state is different from the rest of the RIQHSs 
as it splits into two RIQHSs with a decreasing temperature. 
Such a split is signaled by an $R_{xy}$ deviating from $h/2e^2$ as well as a non-zero
$R_{xx}$ in the vicinity of $\nu=2+2/7$ and it has already been reported in Ref.\cite{xia04}.
The split-off RIQHS is marked as $R2\tilde{a}$ and with a darker shade in Fig.3. 
We note that our data is similar to that in Refs.\cite{eisen02}
in that the $Ria$, $i=2,3$ is the most stable state. Other studies find the $R2c$ state to be the most stable of RIQHSs 
\cite{radu08,dolev11,csa11b,lewis05,csa05,pan08,dean08,umansky09,nuebler10,xia10}.

Each stability boundary shown in Fig.3 can be fitted close to their maxima with a parabolic form $T_c(\nu^*)=T_c(\nu^*_c)-\beta(\nu^*-\nu^*_c)^2$. 
The obtained parameters are listed in Table I. 
$T_c$ obtained from the fit is within 1~mK from the peak temperature obtained from Fig.2. 
The centers $\nu^*_c$ of the RIQHSs in the upper spin branch are in excellent agreement
with the earlier reported values \cite{eisen02}. 
Those of the upper spin branch, however, have not yet been documented
and they differ significantly from those of the lower spin branch. Indeed, $\nu^*_{c,R2\alpha} \neq \nu^*_{c,R3\alpha}$
for $\alpha=a,b,c,$ or $d$, the difference being the largest for the states $a$ and $d$.
Such a difference is not expected from the theory \cite{shibata03,goerbig03} and we think it is due to the
interaction of the electrons in the topmost Landau level with those in the filled lower levels.
Furthermore, we establish that the centers $\nu^*_c$ of RIQHSs in both spin branches
obey particle-hole symmetry, as assumed by the theory \cite{shibata03,goerbig03}. 
In short $\nu^*_{c,Ria}=1-\nu^*_{c,Rid}$ and $\nu^*_{c,Rib}=1-\nu^*_{c,Ric}$ for $i=2,3$, relations which
hold within our measurement error for the filling factor of $\pm 0.003$.

\begin{table}[b]
\caption{Parameters extracted from the $\nu^*$-$T$ diagram. $T_c$ and $\beta$ are in units of mK.}
\begin{ruledtabular}
\begin{tabular}{l c c c c c c c c}
                        & $R2a$ & $R2b$ & $R2c$ & $R2d$ & $R3a$ & $R3b$ & $R3c$ & $R3d$ \\
\hline
$\nu^*_c$               & 0.300 & 0.438 & 0.568 & 0.701 & 0.284 & 0.429 & 0.576 & 0.712 \\
$T_c(\nu^*_c)$          & 53.0  & 37.1  & 45.8  & 38.0  & 46.3  & 32.3  & 36.1  & 33.8  \\
$\beta\times 10^{-4}$   & 10    & 3.9   & 2.4   & 8.5   & 2.1   & 2.0   & 1.6   & 2.3   \\
$\nu^*_{\text{high}}$   & 0.317 & 0.461 & 0.613 & 0.719 & 0.324 & 0.463 & 0.621 & 0.742 \\
$\nu^*_{\text{low}} $   & 0.258 & 0.407 & 0.523 & 0.684 & 0.245 & 0.388 & 0.540 & 0.677 \\
\end{tabular}
\end{ruledtabular}
\end{table}

In contrast to the centers of the RIQHSs, other parameters of the RIQHSs from Table I.
do not obey particle-hole symmetry. These parameters are the maximum onset temperatures $T_c(\nu^*_c)$,
the fit parameter $\beta$ describing the curvature of the stability diagrams near $T_c(\nu^*_c)$,
and the widths $\Delta \nu=\nu^*_{\text{high}}-\nu^*_{\text{low}}$ of the stability regions of the RIQHSs at $T=6.9$~mK.
Indeed, particle-hole symmetry within a spin branch would imply a scaling of $T_c$ with the Coulomb energy $E_C$ 
and, therefore, with $1/\sqrt{\nu}$. Here $E_C=e^2/\epsilon l_B$ and $l_B=\sqrt{\hbar/eB}$ is the magnetic length.
From such a scaling one expects a decreasing $T_c(\nu^*_c)$ with $\nu^*_c$. 
The increasing trend of $T_c$ with $\nu^*$ across $\nu^*=1/2$ shown in Fig.4a 
clearly does not obey such a scaling \cite{pan08} and, 
therefore particle-hole symmetry assumed in current theories \cite{shibata03,goerbig03} is violated. 
The non-monotonic dependence of $T_c$ on $\nu^*_c$ shown in Fig.4a is therefore at odds
with the sequence of the one- and two-electron bubbles suggested \cite{shibata03,goerbig03} 
and could be a consequence of either Landau level mixing, disorder, or 
finite thickness effects. Understanding the origin of this broken symmetry is most likely related to
and, therefore, is expected to impact our understanding of the similar symmetry breaking of the
Pfaffian and anti-Pfaffian construction for the $\nu=5/2$ FQHS \cite{levin07,lee07,wan08,wang09,peterson08,wojs10,bishara09,rezayi11}.

We find that the onset temperatures $T_c(\nu^*_c)$ in the higher spin branch are consistently smaller than  
those in the lower spin branch. We notice, however, a startlingly similar non-monotonic
dependence within each spin branch. A particularly revealing plot is that
of the reduced onset temperatures $T_c(\nu^*_c)/E_C$ against the filling factor $\nu^*_c$. 
As shown in Fig.4b, there is a surprizingly good collapse of $T_c(\nu^*_c)/E_C$ for the
different spin branches. This collapse shows that Coulomb
interactions play a central role in the formation of the RIQHSs in the SLL and provides
a first direct evidence that these states reflect collective 
behavior of the electrons rather than single particle localization.   
The lack of collapse of $T_c(\nu^*_c)/\hbar\omega_C$ shown in Fig.4c 
means that $T_c(\nu^*_c)$ does not scale with the cyclotron energy $\hbar\omega_C$. 

\begin{figure}[t]
 \includegraphics[width=1\columnwidth]{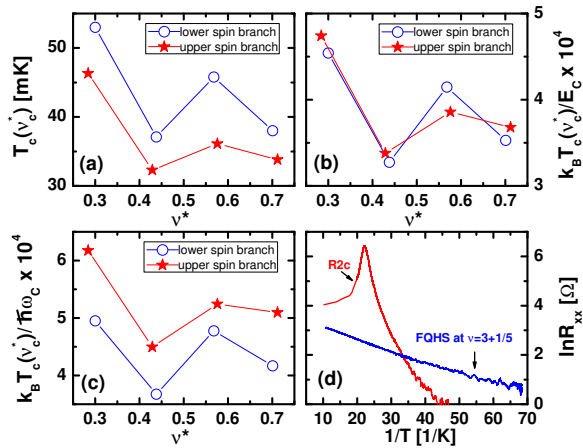}
 \caption{ The variation  with the filling factor $\nu^*_c$ of the onset temperatures $T_c(\nu^*_c)$ 
 at the center in units of mK (panel a), Coulomb energy (panel b), and cyclotron energy (panel c)
 of the eight observed RIQHSs in the SLL. Panel d is an Arrhenius plot for $R2c$ and for the $\nu=3+1/5$ FQHS. 
 Lines are guides to the eye.
 \label{Fig4}}
 \end{figure}  

In a recent study an activated dependence of $R_{xx}(T)$ is found for the $R2c$ state \cite{nuebler10}.
In our sample we find a significant deviation from such a dependence and, as a consequence, 
the definition of an activation energy
is no longer possible. Fig.4d shows such a plot, together with the activated resistance of a FQHS measured 
in order to rule out thermometry artifacts. Our data suggest that non-activated behavior might be an inherent property of the RIQHSs. 

In summary we find that
the scaling of the onset temperatures in different spin branches with the Coulomb energy
provides a direct experimental evidence for the collective nature of the RIQHSs of the SLL. 
The stability diagram we report in the $\nu^*$-$T$ plane reveals several quantitative disagreements
with the existing theories such as the lack of paricle-hole symmetry of the onset temperatures
within one spin branch. 

The work at Purdue was supported by the DOE grant DE-SC0006671. 
L.N.P. and K.W.W. were supported by the Princeton NSF-MRSEC and the Moore Foundation. 
We benefited from discussions with G. Giuliani and S. Das Sarma.

\end{document}